\documentclass[runningheads]{llncs}
\usepackage{graphicx, amsmath}
\usepackage{multirow}
\usepackage{pdfpages}
\usepackage{amssymb}
\usepackage{mathrsfs}
\usepackage{cite}
\usepackage{color}
\usepackage[colorlinks=true, linkcolor=red, citecolor=green]{hyperref}
\usepackage{enumitem}
\usepackage{mwe} 
\begin{document}
\title{Self-supervised arbitrary scale super-resolution framework for anisotropic MRI}
%
%
\author{Haonan Zhang\inst{1, *} \and
Yuhan Zhang\inst{1, *} \and
Qing Wu\inst{1} \and
Jiangjie Wu\inst{1} \and
Zhiming Zhen\inst{2} \and
Feng Shi\inst{3} \and
Jianmin Yuan\inst{4} \and
Hongjiang Wei\inst{5} \and
Chen Liu\inst{2},\inst{6} \and
Yuyao Zhang \inst{1},\inst{6}}
\authorrunning{F. Haonan Zhang et al.}
%
\institute{School of Information Science and Technology, ShanghaiTech University, Shanghai, China \and
Department of Radiology, Southwest Hospital, Army Medical University \\ (Third Military Medical University), Chongqing, 400038, China \and
Shanghai United Imaging Intelligence Co., Ltd., Shanghai, China \and
Central Research Institute, UIH Group, Shanghai, China \and
School of Biomedical Engineering, Shanghai Jiao Tong University, Shanghai, China \and
\texttt{Co-corresponding Authors}}
\maketitle              
\begin{abstract}
In this paper, we propose an efficient self-supervised arbitrary-scale super-resolution (SR) framework to reconstruct isotropic magnetic resonance (MR) images from anisotropic MRI inputs without involving external training data. The proposed framework builds a training dataset using ”in-the-wild” anisotropic MR volumes with arbitrary image resolution. We then formulate the 3D volume SR task as a SR problem for 2D image slices. The anisotropic volume’s high-resolution (HR) plane is used to build the HR-LR image pairs for model training. We further adapt the implicit neural representation (INR) network to implement the 2D arbitrary-scale image SR model. Finally, we leverage the well-trained proposed model to up-sample the 2D LR plane extracted from the anisotropic MR volumes to their HR views. The isotropic MR volumes thus can be reconstructed by stacking and averaging the generated HR slices. Our proposed framework has two major advantages: (1) It only involves the arbitrary-resolution anisotropic MR volumes, which greatly improves the model practicality in real MR imaging scenarios (e.g., clinical brain image acquisition); (2) The INR-based SR model enables arbitrary-scale image SR from the arbitrary-resolution input image, which significantly improves model training efficiency. We perform experiments on a simulated public adult brain dataset and a real collected 7T brain dataset. The results indicate that our current framework greatly outperforms two well-known self-supervised models for anisotropic MR image SR tasks.

\end{abstract}
\section{Introduction}
\par Magnetic Resonance Imaging (MRI) is an essential medical imaging technology. However, isotropic 3D High-Resolution (HR) MR images are difficult to acquire due to the trade-off problem among image resolution, Signal-Noise-Ratio (SNR), and scanning time \cite{jia2017new}. A common compromise is to scan anisotropic MR images to improve SNR and shorten the scanning time. Unfortunately, the measured anisotropic MR volumes have high in-plane resolutions but low through-plane resolutions, which results in the loss of high-frequency image details in the through-plane view and may hinder the following works (e.g., lesion detection and segmentation). Thus improving the through-plane resolution to reconstruct the isotropic MR images is an urgent need for clinical medical diagnosis and research.
\begin{figure}[t]
    \centering
    \includegraphics[width=0.38\textwidth]{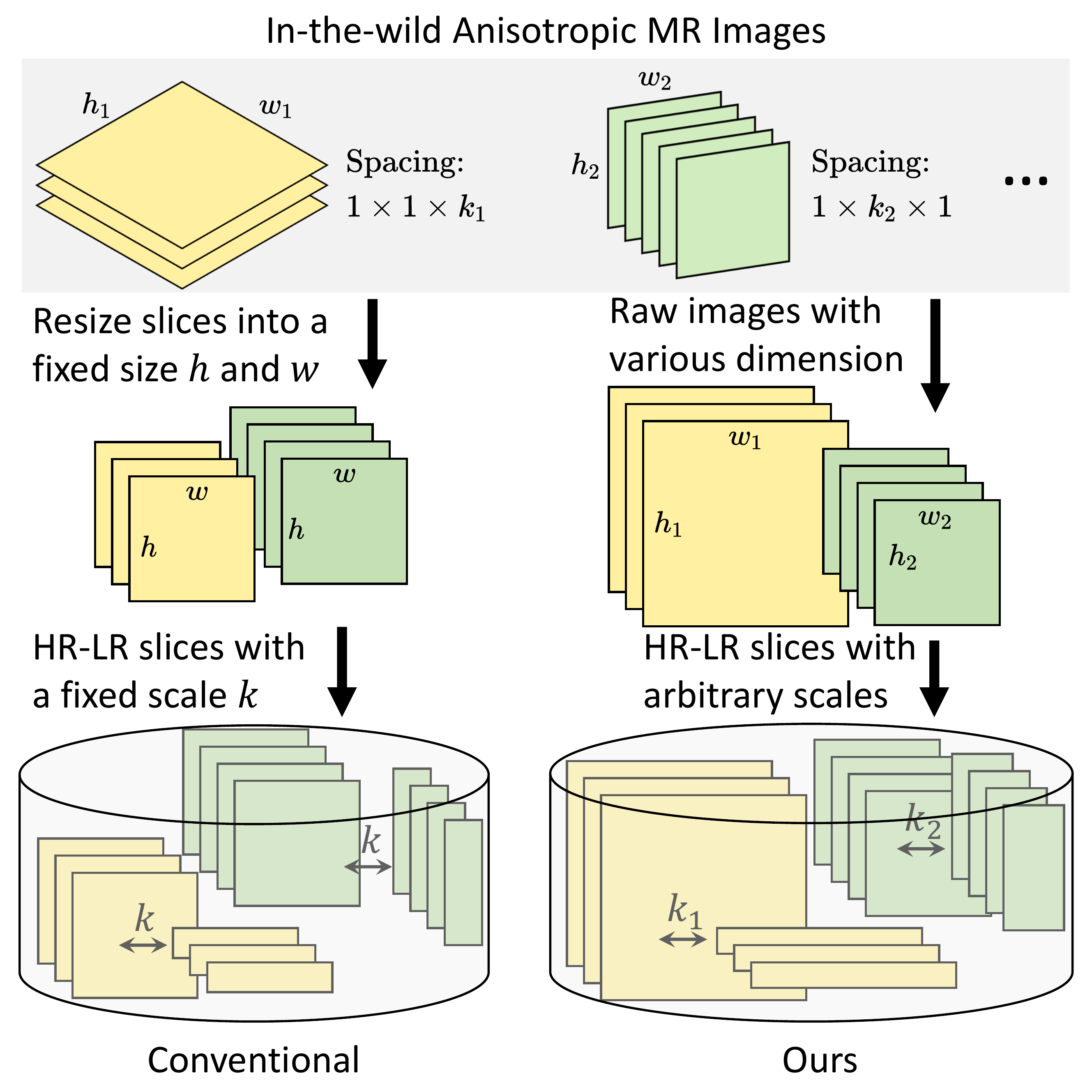}
    \caption{Pipelines of building SR dataset from in-the-wild multiple anisotropic MR images. The conventional pipeline \cite{smore,weigert2017isotropic} is to resize the raw slices into a fixed size and simulate the paired HR-LR slices of a fixed scale. While we aim to keep the slices at their raw resolutions and build the paired HR-LR slices of arbitrary scales.}
    \label{fig:dataset}
\end{figure}
\par Currently, Convolutional Neural Networks (CNNs) are mainstream solutions for isotropic MR image Super-Resolution (SR) reconstruction. Most CNNs-based SR methods \cite{pham2017brain,chen2018brain,du2020super,chen2018efficient} learn inverse mappings from anisotropic inputs to isotropic outputs by training CNNs over external datasets. They produce robust and excellent SR performance benefiting from the data-driven priors. However, the fixed inverse mappings learned by CNNs suffer from severe performance drops when the up-sampling scale changes \cite{kim2016accurate,hu2019meta,chen2021learning}. Therefore, an independent CNN needs to be trained in practice for each up-sampling scale. This scale-specific learning paradigm is thus extremely resource-intensive.
\par Recently, several arbitrary-scale MR image SR methods \cite{wu2021arbitrary,wang2022arbitrary,van2022scale} based on Implicit Neural Representation (INR) have emerged. INR is a signal representation based on neural networks. The signal is formulated as a continuous function of spatial coordinates and is approximated by a Multi-Layer Perceptron (MLP). Due to the continuous representation provided by INR, the single well-trained INR-based model theoretically can handle the SR tasks of arbitrary scales and thus significantly reduce resource consumption. However, the INR-based SR methods \cite{wu2021arbitrary,wang2022arbitrary,van2022scale} are almost supervised; thus, a large-scale external dataset is required for training.  Moreover, scanning the isotropic 3D HR MR volumes is often time-consuming and expensive due to the trade-off problem, which significantly limits model performance and practicality.
\begin{figure}[t]
    \centering
    \includegraphics[width=0.48\textwidth]{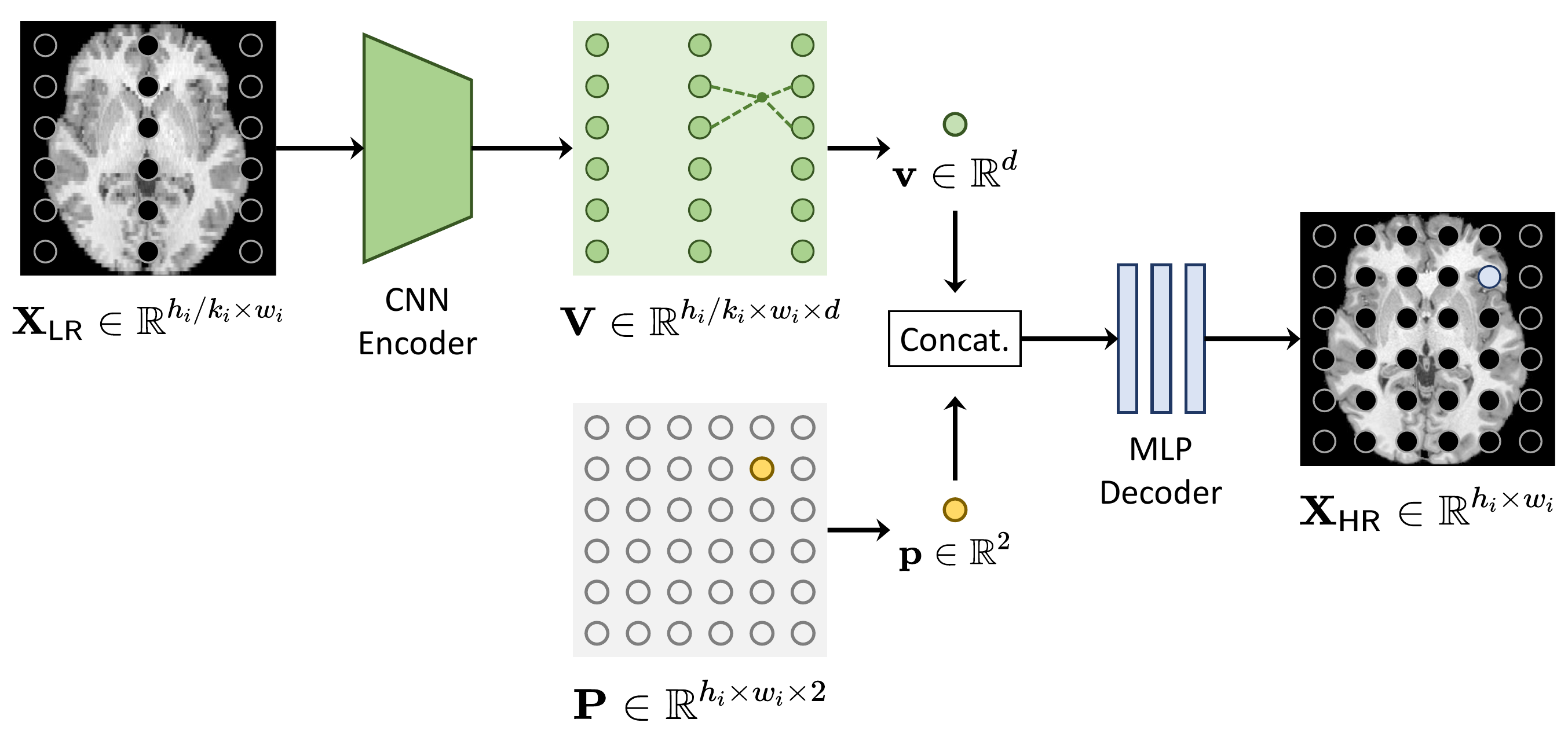}
    \caption{Pipeline of our arbitrary-scale SR model.}
    \label{fig:method}
\end{figure}
\par In this work, we propose a self-supervised arbitrary-scale SR framework, in which an arbitrary-scale SR model is trained based on the multiple anisotropic MR images (Note the multiple images refer to ”in-the-wild“ MR volumes with arbitrary image resolution, instead of forcibly requiring anisotropic images acquired from orthogonal views.). Our single well-trained SR model can reconstruct the corresponding isotropic HR MR volumes from the multiple anisotropic inputs without involving any external training data.
\par To sum up, there are two major contributions in our proposed framework: (1) We propose a universal pipeline for constructing an arbitrary-resolution and arbitrary-scale SR dataset from the anisotropic MR volumes, which releases the need for external isotropic 3D HR MR images (Sec. \ref{sec:data_construction}); (2) We propose an arbitrary-scale SR model based on INR for 2D MR slices and train it on the built SR dataset (Sec. \ref{sec:model}). After model training, we utilize the SR model to up-sample the 2D LR slices extracted from the anisotropic MR images along their HR views. The final isotropic volumes thus can be reconstructed by stacking and averaging the generated HR slices (Sec. \ref{sec.test}).
\par To evaluate the proposed framework, we conduct comparison experiments on two MR image datasets (including a simulation adult dataset and a real collection fetus dataset). The results demonstrate that the proposed framework outperforms the two self-supervised methods for isotropic 3D HR MR image SR tasks while using a single-trained model to handle SR tasks of arbitrary scales.
\label{sec:intro}
\section{Proposed Framework}
\subsection{Arbitrary-Resolution and -Scale SR Dataset Construction}
\label{sec:data_construction}
\par In 2017, \cite{smore,weigert2017isotropic} proposed a pipeline for building an SR training dataset from in-the-wild multiple anisotropic MR images. The main idea is to consider the 2D HR slices along the LR view as GT images and then perform down-sampling on the HR slices to simulate the paired 2D HR-LR slices training data. As shown in Figure \ref{fig:dataset} (Left), the conventional pipeline proposed by \cite{smore,weigert2017isotropic} first resizes the raw HR slices into a fixed size. Then, down-sampling the HR slices by a fixed scale to simulate the corresponding LR slices. However, there are two limitations: (1) Resizing operator will cause a loss of the multi-scale information provided by the raw arbitrary-resolution HR slices; (2) The paired HR-LR slices of a fixed scale only can be used to train scale-specific SR models.
\par To this end, we propose constructing an arbitrary-resolution and scale SR dataset. The pipeline is demonstrated in Figure \ref{fig:dataset} (Right). Given the measured multiple anisotropic MR volumes, we directly extract raw 2D slices along their LR views as the GT HR slices without using any resizing operator, benefiting the preserving of the original multi-resolution image information. For example, 35 slices of $200\times200$ size will be extracted for an anisotropic volume of $35\times200\times200$ size. Then, we down-sample the HR slices to simulate the corresponding 2D LR slices. Instead of the fixed scale in the conventional pipeline \cite{smore,weigert2017isotropic}, the down-sampling scales are set as the ratios between in-plane spacing and through-plan spacing in the original 3D anisotropic volumes. Finally, the generated paired HR-LR slices have arbitrary scales due to the anisotropic inputs of various spacings (i.e., voxel sizes). Therefore, our build dataset can be used for training arbitrary-scale SR models.
\begin{figure}[t]
    \centering
    \includegraphics[width=0.48\textwidth]{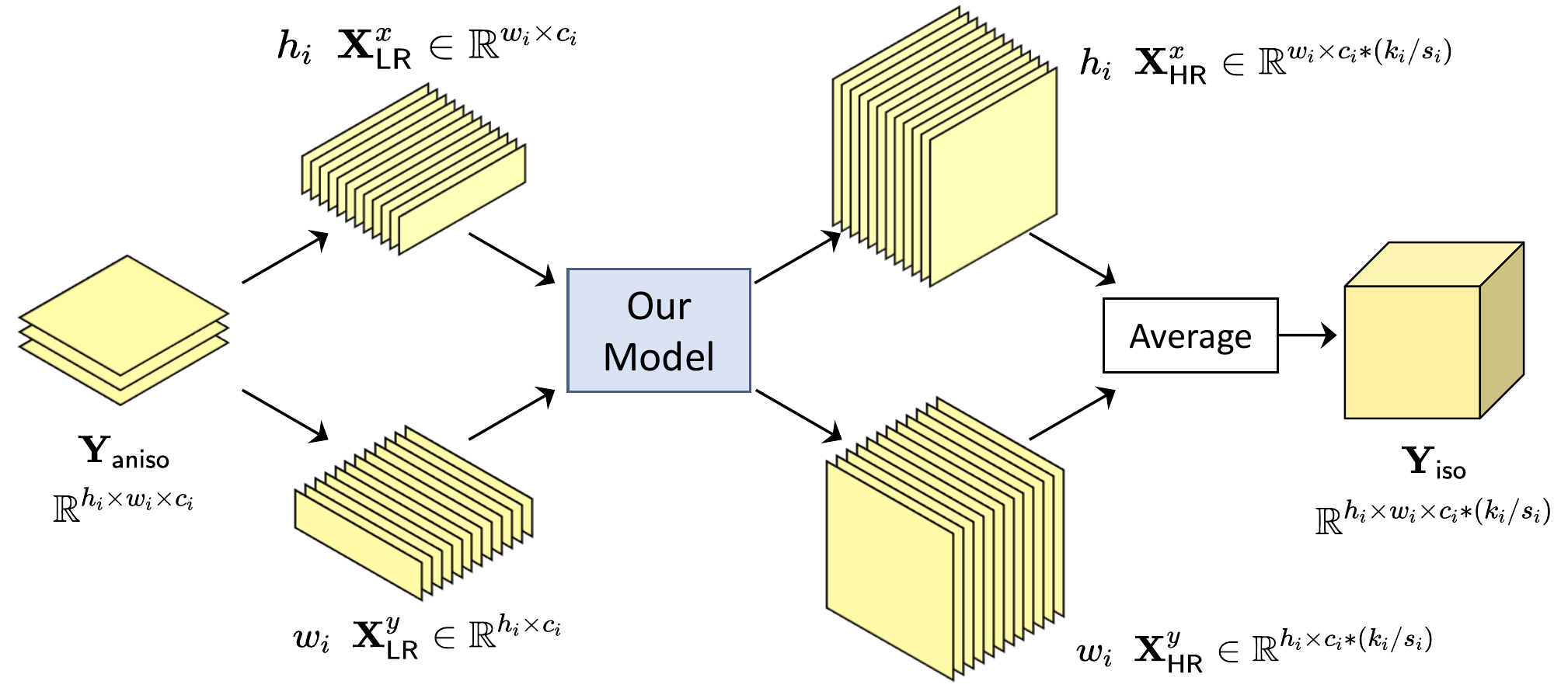}
    \caption{Pipeline of an isotropic 3D HR volume reconstruction by our arbitrary-scale SR model.}
    \label{fig:test}
\end{figure}
\begin{table}[t]
    \centering
    \resizebox{0.45\textwidth}{!}{
    \begin{tabular}{c|c|c|c}
        \textbf{Dataset} & \textbf{Image}  & \textbf{Spacing (mm)}          & \textbf{Image Size}  \\ \hline
        \multirow{5}{*}{\textbf{HCP-1200} \cite{hcp}} 
        & Sub \#1   & $1.40\times0.70\times0.70$       & $130\times260\times260$ \\ 
        & Sub \#2   & $0.70\times1.75\times0.70$      & $260\times104\times260$ \\  
        & Sub \#3   & $0.70\times0.70\times2.10$       & $260\times260\times87$  \\ 
        & Sub \#4   & $2.45\times0.70\times0.70$      & $74\times260\times260$  \\ 
        & Sub \#5   & $0.70\times2.80\times0.70$       & $260\times65\times260$  \\ \hline
        \multirow{2}{*}{\textbf{7T Adult Brain}}    
        & Adult \#1 & $2.00\times0.41\times0.41$         & $80\times544\times544$  \\ 
        & Adult \#2 & $3.00\times0.73\times0.73$ & $49\times288\times288$  \\
    \end{tabular}}
    \caption{Details of the two datasets used in our experiments.}
    \label{tab:data}
\end{table}
\begin{figure*}[t]
    \centering
    \includegraphics[width=0.80\textwidth]{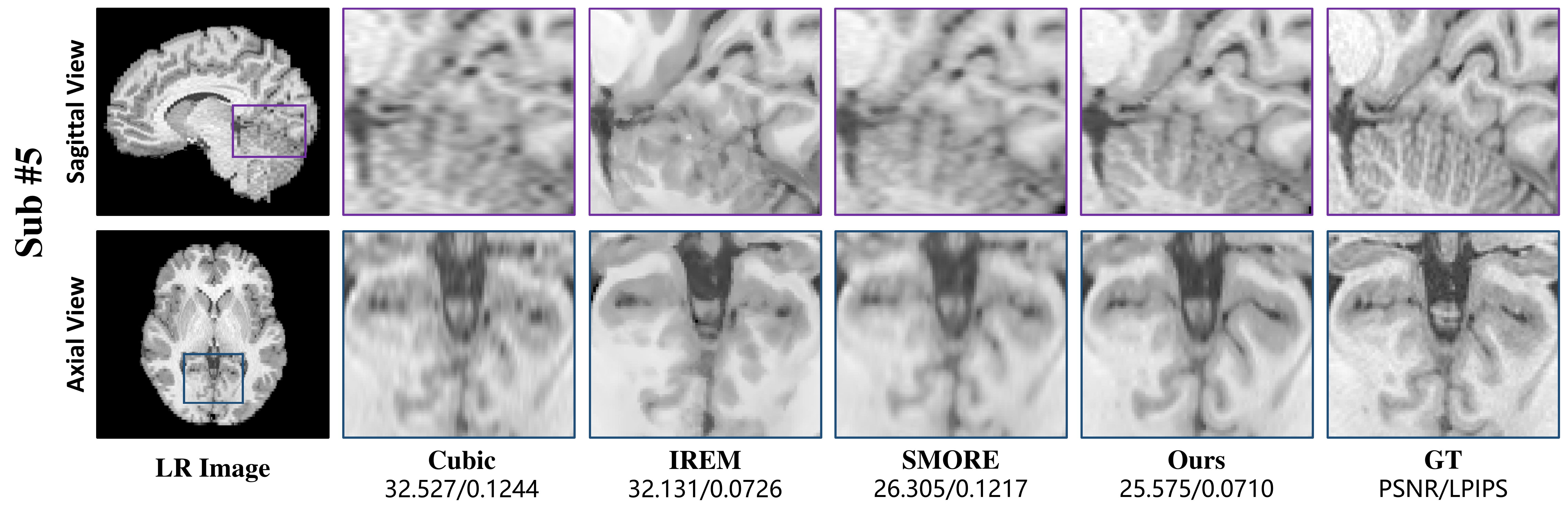}
    \caption{Qualitative results of all compared methods on Sub \#5 of the simulated HCP-1200 data \cite{hcp} for $4\times$ SR at coronal view.}
    \label{fig_simu}
\end{figure*}
\begin{table*}[t]
    \centering
    \resizebox{0.76\textwidth}{!}{
    \begin{tabular}{c|c|c|c|c}
       \textbf{Scale \& View}  &\textbf{Cubic}& \textbf{IREM} \cite{wu2021irem}  &\textbf{SMORE} \cite{smore}&  \textbf{Ours}\\ \hline
        $2\times$ \& sagittal& 
        $39.616/0.9760/0.0297$& 
        $37.632/0.9570/0.0783$&
        $41.514/0.9857/0.0259$& 
        $38.928/0.9734/0.0269$\\
        $2.5\times$ \& coronal& $32.254/0.9499/0.0781$& $31.345/0.9345/0.1019$& $27.903/0.9061/0.0778$& $31.038/0.9510/0.0518$\\
        $3\times$ \& axial&
        $36.764/0.9702/0.0718$& 
        $34.889/0.9523/0.0632$&
        $32.749/0.9470/0.0757$& 
        $32.255/0.9719/0.0286$\\
        $3.5\times$ \& sagittal& $25.568/0.8904/0.1398$& $25.370/0.8813/0.1004$& $21.320/0.8541/0.1436$& $25.064/0.8857/0.0781$\\ 
        $4\times$ \& coronal&
        $32.527/0.9244/0.1244$& 
        $32.131/0.9181/0.0726$& 
        $26.305/0.8619/0.1217$& 
        $25.575/0.9144/0.0710$\\ \hline
        \textbf{Mean} &
        $33.346/0.9422/0.0888$& 
        $32.273/0.9286/0.0833$&
        $29.958/0.9110/0.0889$& 
        $30.572/0.9393/0.0513$ \\
    \end{tabular}}
    \caption{Quantitative results (PSNR/SSIM/LPIPS) of all compared methods on the simulated HCP-1200 data \cite{hcp}.}
    \label{tab:com}
\end{table*}
\subsection{Arbitrary-Scale SR Model for 2D MR Slices}
\label{sec:model}
\par Inspired by \cite{wu2021arbitrary}, we propose an arbitrary-scale SR model for 2D MR slices. Given any pair of LR-HR images $\{\mathbf{X}_{\mathsf{LR}}\in\mathbb{R}^{{h_i}/{k_i}\times w_i}, \mathbf{X}_{\mathsf{HR}}\in\mathbb{R}^{h_i\times w_i}\}$ from our arbitrary-resolution and -scale dataset, where $k_i$ represents their up-sampling scale, we leverage a continuous function to represent them as below:
\begin{equation}
    f_{\theta}: \left(\mathbf{p}, \mathbf{v}\left(\mathbf{p}\right)\right) \rightarrow \mathbf{X}(\mathbf{p}),
    \label{eq1}
\end{equation}
where $\mathbf{p}=(x,y)\in\mathbb{R}^2$ is any spatial coordinate in a normalized 2D Cartesian coordinates $[-1, 1]\times[-1, 1]$. $\mathbf{X}(\mathbf{p})\in\mathbb{R}$ and $\mathbf{v}(\mathbf{p})\in\mathbb{R}^d$ denote the intensity value and semantic representation at the position $\mathbf{p}$ in the image $\mathbf{X}$ respectively. The vector $\mathbf{v}(\mathbf{p})$ is used for a conditional input such that the function $f_\theta$ can be shared for different MR slices. Therefore, the paired HR-LR image $\{\mathbf{X}_{\mathsf{LR}}, \mathbf{X}_{\mathsf{HR}}\}$ can be considered as the discrete explicit representation of the continuous function $f_\theta$ at different sampling spacing.
\par Based on the definition above, the arbitrary-scale SR task is converted to learn the continuous function $f_\theta$. Figure \ref{fig:method} shows the pipeline. We first employ a CNN encoder $\mathcal{C}_\phi$ to map the LR image $\mathbf{X}_{\mathsf{LR}}$ to a feature map $\mathbf{V}\in\mathbb{R}^{{h_i}/{k_i}\times w_i\times d}$, where an element $\mathbf{v}$ is the local semantic representation of the input image $\mathbf{X}_{\mathsf{LR}}$. Then, for a query coordinate $\mathbf{p}$ in HR grid $\mathbf{P}\in\mathbb{R}^{h_i\times w_i\times 2}$ built on the HR image $\mathbf{X}_{\mathsf{HR}}$, we generate the corresponding feature vector $\textbf{v}(\mathbf{p})$ by performing linear interpolation on the feature map $\mathbf{V}$. Then, the vector $\mathbf{v}(\mathbf{p})$ as a conditional input that concatenated with the coordinate $\mathbf{p}$ is feed into a Multi-Layer Perceptron (MLP) decoder $\mathcal{M}_\phi$ to predict the HR intensity value $\hat{\mathbf{X}}_{\mathsf{HR}}(\mathbf{p})$. Finally, we simultaneously optimize the CNN encoder and MLP decoder to learn the function by minimizing the objective as below:
\begin{align}
\mathbf{\phi}^*, \mathbf{\psi}^* &= \mathop{\arg\min}\limits_{\phi, \psi}\ \mathcal{L}(\mathbf{X}_{\mathsf{HR}}(\mathbf{p}), \hat{\mathbf{X}}_{\mathsf{HR}}(\mathbf{p})),\\
\hat{\mathbf{X}}_{\mathsf{HR}}(\mathbf{p}) &= \mathcal{M}_\psi(\mathbf{p},
\mathbf{v}(\mathbf{p})),\\
\mathbf{v}(\mathbf{p}) &= \mathsf{Linear\ Inter.}(\mathbf{V}), \\
\mathbf{V} &= \mathcal{C}_\phi(\mathbf{X}_{\mathsf{LR}}),
\end{align}
where the loss function $\mathcal{L}$ is implemented by $\ell_1$ norm. We use Adam optimizer to train the model, and its hyper-parameters are default. The initial learning rate is 1e-4, and the training epoch is 800. The best model is saved by checkpoints during the training process.
\subsection{Isotropic 3D HR Volumes Reconstruction}
\label{sec.test}
\par Figure \ref{fig:test} illustrates the pipeline of an isotropic 3D HR volume reconstruction by our well-trained model. Let $\mathbf{Y}_{\mathsf{aniso}}\in\mathbb{R}^{h_i\times w_i \times c_i}$ of a spacing $s_i\times s_i\times k_i$ ($s_i<k_i$) denote any one of the anisotropic input volumes, our purpose is to recover the corresponding isotropic image $\mathbf{Y}_{\mathsf{iso}}$ of a spacing $s_i\times s_i\times s_i$ by using the well-trained ArSSR2D model. To this end, we first extract $h_i$ LR slices $\mathbf{X}_{\mathsf{LR}}^x\in\mathbb{R}^{w_i\times c_i}$ and $w_i$ LR slices $\mathbf{X}_{\mathsf{LR}}^y\in\mathbb{R}^{h_i\times c_i}$ from $\mathbf{Y}_{\mathsf{aniso}}$ along two HR views, respectively. Then, these LR slices are fed our SR model to generate their HR versions, i.e., $h_i$ $\mathbf{X}_{\mathsf{HR}}^x\in\mathbb{R}^{w_i\times c_i*(k_i/s_i)}$ and $w_i$ $\mathbf{X}_{\mathsf{HR}}^y\in\mathbb{R}^{h_i\times c_i*(k_i/s_i)}$. Finally, we can reconstruct the isotropic volume $\mathbf{Y}_{\mathsf{iso}}\in\mathbb{R}^{h_i\times w_i \times c_i*(k_i/s_i)}$ by stacking and averaging the generated HR slices. 
\begin{figure*}[t]
    \centering
    \includegraphics[width=0.89\textwidth]{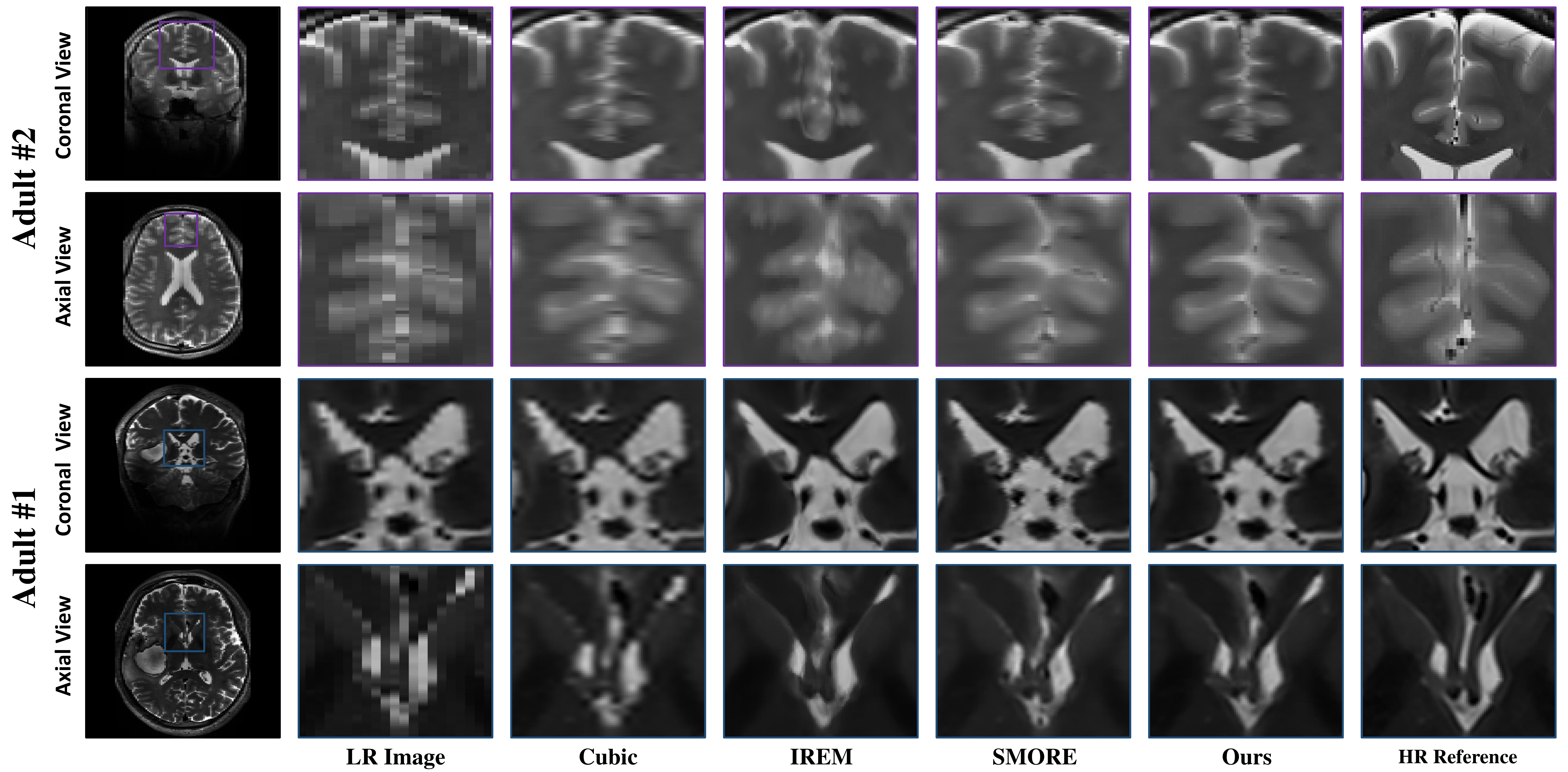}
    \caption{Qualitative results of all compared methods on the two samples of the real 7T adult brain data for $4.86\times$ and $4.11\times$ SR at sagittal view.}
    \label{fig_real}
\end{figure*}
\section{Experiments}
\label{sec:experiments}
\subsection{Experimental Setup}
\label{ssec:Experimental Setup}
\par\noindent\textbf{Simulated HCP-1200 Data:} HCP-1200 dataset \cite{hcp} is a large-scale public adult brain MR image dataset consisting of 1113 3D isotropic HR multi-modal brain MR images. In our experiments, we use five isotropic HR T1-weighted MR images of the HCP-1200 dataset \cite{hcp} as Ground Truth (GT) images. To simulate anisotropic LR input images, we then respectively downsample the five GT images by the scales $\{2, 2.5, 3, 3.5, 4\}$ along the $\{x, y, z, x, y\}$ views, where $x$, $y$, and $z$ denote respectively sagittal, coronal, and axial views. The GT images are only used for final model evaluation, while all the compared models do not see them during the model training.
\par\noindent\textbf{Real 7T Adult Brain Data:} We scan two T2-weighted anisotropic brain MR images from two adults \textit{in vitro} on a 7T Siemens MR scanner. The spacing of the raw images are $2\times0.4118\times0.4118$ mm and $3\times0.2431\times0.2431$ mm respectively. The size of the raw images are $80\times544\times544$ and $49\times864\times864$ respectively. To reduce computational consumption while maintaining image content as much as possible, we decrease the latter size to $49\times288\times288$ by cropping.
\par\noindent\textbf{Compared Methods:} We compare our model with three SR methods: (1) Cubic interpolation, a mathematical method based on cubic polynomials; (2) IREM, a self-supervised SR method based on INR. (3) SMORE \cite{smore}, a deep learning method based on self super-resolution (SRR) \cite{srr}. Here Cubic interpolation is based on the scikit-image library of Python, while IREM \cite{wu2021irem} and SMORE \cite{smore} are reproduced following the original papers.
\par\noindent\textbf{Evaluation Metrics:} PSNR and SSIM \cite{ssim}, two widely-used image quality evaluation metrics, are computed. Moreover, we also calculate LPIPS \cite{lpips}, a popular deep-learning-based similarity metric for a more comprehensive evaluation. The slice-by-slice strategy is employed to calculate LPIPS \cite{lpips} since it is designed for 2D images. Specifically, we first extract the most central 30 2D slices from SR volumes along three orthogonal views (i.e., ten images per view). Then, we calculate LPIPS \cite{lpips} based on the 2D slices and average them to obtain the final scores.
\subsection{Results on Simulated HCP-1200 Data}
\label{ssec:Results On Simulated Data}
\par We compare our proposed method with the three baselines based on the five simulated anisotropic MR volumes (details in Table \ref{tab:data}) of the HCP-1200 dataset\cite{hcp}. Specifically, we apply the four methods to upsample the five anisotropic adult brain images to generate the corresponding isotropic volumes. It is worth noting that IREM \cite{wu2021irem} and SMORE \cite{smore} are independently trained for each input volume, while our single well-trained model is shared for the five different volumes.
\par Figure \ref{fig_simu} shows the qualitative results on Sub \#5 for the $4\times$ SR task at the coronal view. Cubic interpolation produces a very blurry SR image. IREM \cite{wu2021irem} yields an overly smooth image. The SR result from SMORE \cite{smore} is excellent in image structure but unclear in image details. In comparison, the SR image of our method is closest to GT in both image sharpness and consistency. Moreover, Figure \ref{fig_simu} shows an interesting result: although the SR result by cubic interpolation is worst from the visualization, it obtains the highest PSNR (32.527 dB). As claimed in \cite{wang2020enhanced,chen2018efficient}, we hold that, the PSNR metric is defined on pixel-by-pixel distance and thus is limited for evaluating SR tasks. Instead, LPIPS \cite{lpips} is measured based on semantic representations extracted by a pre-trained deep learning model. For Sub \#5, our model obtains the best performance (0.071) in terms of LPIPS, which is consistent with our visual observation. We also report the quantitative results in Table \ref{tab:com}. We observe that cubic interpolation yields the highest PSNR and SSIM (33.346 dB and 0.9422), while our proposed method obtains the best performance (0.0515) in LPIPS.
\subsection{Results on Real 7T Adult Brain Data}
\label{ssec:Result On Real Data}
\par Although our proposed framework yields significant improvements on the simulated HCP-1200 data \cite{hcp} compared with the three baselines, it is more important to show how well the models perform on the real acquired anisotropic MR volumes. Therefore, we also conducted a comparison experiment on two real measured anisotropic adult brain volumes (details in Table \ref{tab:data}). Here we only report the qualitative results since the isotropic GT HR images are not acquired. 
\par As shown in Figure \ref{fig_real}, the SR results of cubic interpolation suffer from severe blocking artifacts due to its poor de-aliasing ability. While IREM \cite{wu2021irem} produces sharp but overly smooth MR images, where many high-frequency details are lost. In comparison, the SR results of SMORE \cite{smore} and our proposed model recover more image details. Moreover, the SR results of our model are better in terms of anatomical consistency.

\section{Conclusion}
\label{sec:conclusion}
\par This work presents a self-supervised deep-learning framework to reconstruct the isotropic volumes from multiple anisotropic MR images when without any extra data. In the proposed framework, we first construct an arbitrary-scale SR dataset based on the anisotropic volumes, then train an arbitrary-scale SR model on our built dataset. The well-trained single SR model can be used to reconstruct the HR isotropic MR images. The experimental results on the simulated HCP-1200 \cite{hcp} data and the real 7T adult brain data indicate that our proposed method can recover excellent HR isotropic MR images. Our proposed self-supervised framework has great application potential for improving MR image quality.
\section{Compliance with ethical standards}
\label{sec:ethics}
This study was performed in line with the principles of the Declaration of Helsinki. Approval was granted by the Ethics Committee of the Southwest hospital, Chongqing, China.
\section{Acknowledgments}
\label{sec:acknowledgments}
This work is supported by the National Natural Science Foundation of China (No. 62071299, 61901256, 91949120).

\bibliographystyle{IEEEbib.bst}
\bibliography{refs}

\end{document}